\documentclass[3p,times]{elsarticle}

\usepackage{ecrc}

\volume{00}

\firstpage{1}

\journalname{Nuclear Physics A}

\runauth{}

\jid{nupha}

\usepackage{amssymb}

\usepackage[figuresright]{rotating}
\usepackage{graphicx}
\usepackage{epstopdf}

\begin{document}
\def\Journal#1#2#3#4{{#1} {\bf #2}, #3 (#4)}

\def\NCA{Nuovo Cimento}
\def\NIM{Nucl. Instr. Meth.}
\def\NIMA{{Nucl. Instr. Meth.} A}
\def\NPB{{Nucl. Phys.} B}
\def\NPA{{Nucl. Phys.} A}
\def\PLB{{Phys. Lett.}  B}
\def\PRL{Phys. Rev. Lett.}
\def\PRC{{Phys. Rev.} C}
\def\PRD{{Phys. Rev.} D}
\def\ZPC{{Z. Phys.} C}
\def\JPG{{J. Phys.} G}
\def\CPC{Comput. Phys. Commun.}
\def\EPJ{{Eur. Phys. J.} C}
\def\PR{Phys. Rept.}
\def\JHEP{JHEP}
\begin{frontmatter}



\dochead{}

\title{The di-lepton physics program at STAR}


\author{Lijuan Ruan for the STAR Collaboration}

\address{Physics Department, Brookhaven National laboratory, Upton NY 11973}
\ead{ruanlj@rcf.rhic.bnl.gov, ruan@bnl.gov}
\begin{abstract}
The recent results on di-electron production in $p+p$ and Au+Au
collisions at $\sqrt{s_{_{NN}}} = 200$ GeV are presented. The
cocktail simulations of di-electrons from light and heavy flavor
hadron decays are reported and compared with data. The
perspectives for di-lepton measurements in lower energy Au+Au
collisions and with future detector upgrades are discussed.
\end{abstract}

\begin{keyword}
di-electron continuum, cocktail simulation, low-mass enhancement,
QGP thermal radiation, $\mu-e$ correlation


\end{keyword}

\end{frontmatter}



\section{Introduction}
Ultra-relativistic heavy ion collisions provide a unique
environment to study the properties of strongly interacting matter
at high temperature and high energy density~\cite{starwhitepaper}.
One of the crucial probes of this strongly interacting matter are
di-lepton measurements in the low and intermediate mass region.
Di-leptons are not affected by the strong interaction once
produced, therefore they can probe the whole evolution of the
collision.

In the low invariant mass range of produced lepton pairs
($M_{ll}\!<\!1.1$ GeV/$c^{2}$), we can study vector meson
in-medium properties through their di-lepton decays, where
modifications of mass and width of the spectral functions observed
may relate to the possibility of chiral symmetry
restoration~\cite{dilepton,dileptonII}. At the SPS, the low mass
di-lepton enhancement in the CERES $e^+e^-$ data ~\cite{ceres} and
in the NA60 $\mu^+\mu^-$ data~\cite{na60} requires substantial
medium effects on the $\rho$-meson spectral function. The precise
NA60 measurement of the low mass enhancement provides a decisive
discrimination between a dropping-mass scenario~\cite{dropmass}
and a massively broadened spectral function~\cite{massbroaden}.
The latter one was found to be able to consistently describe the
data.

The di-lepton spectra in the intermediate mass range
($1.1\!<M_{ll}\!<\!3.0$ GeV/$c^{2}$) are directly related to
thermal radiation of the Quark-Gluon Plasma
(QGP)~\cite{dilepton,dileptonII}. However, contributions from
other sources have to be measured experimentally. Such
contributions include background pairs from correlated open heavy
flavor decays, which produce a pair of electrons or muons from the
semileptonic decay of a pair of open charm or bottom hadrons
($c\bar{c}\rightarrow l^{+}l^{-}X$ or $b\bar{b}\rightarrow
l^{+}l^{-}X$).

Anisotropic flow, an anisotropy in the particle production
relative to the reaction plane, leads to correlations among
particles~\cite{Art:98}. The elliptic flow $v_2$ is the second
harmonic of the azimuthal distribution of particles with respect
to the reaction plane. $v_2$ has been measured for direct photons
and found to be substantial in the transverse momentum range
$1\!<\!p_{T}\!<\!4$ GeV/$c$ in central 0-20\% Au+Au collisions at
$\sqrt{s_{_{NN}}} = 200$ GeV~\cite{photonv2}. In the same $p_T$
region, PHENIX measured direct photon yields and found an excess
of direct photon yield in 0-20\% Au+Au over $p+p$, exponential in
$p_T$~\cite{thermalphoton}. Model calculations~\cite{Rupa:09} for
QGP thermal photons in this kinematic region significantly
under-predict the observed $v_2$ while if a significant
contribution from the hadronic sources at later stages is added,
the excess of the spectra and the observed $v_2$ at
$1\!<\!p_{T}\!<\!4$ GeV/$c$ are described reasonably
well~\cite{rapp:11}. It has been proposed that di-lepton $v_2$
measurements will provide another independent way to study medium
properties since di-leptons provide two independent kinematic
parameters: mass and $p_T$. Specifically, $v_2$ as a function of
$p_T$ in different mass regions will enable us to probe the
properties of medium from a hadron-gas dominated to a QGP
dominated scenario~\cite{Gale:07}.

At STAR, the newly installed Time-of-Flight detector (TOF) offers
large acceptance and high efficiency~\cite{startof}. The TOF,
combined with measurements of ionization energy loss (dE/dx) from
the Time Projection Chamber (TPC)~\cite{startpc,bichsel,pidpp08},
enables electron identification with high purity for
$0.2\!<\!p_{T}\!<\!3$ GeV/$c$~\cite{pidNIMA,tofPID,starelectron}.
In this article we present the di-electron mass spectra in $p+p$
and Au+Au collisions at $\sqrt{s_{_{NN}}} = 200$ GeV.  The
elliptic flow $v_2$ measurements are also reported in 200 GeV
Au+Au collisions. Future capabilities for di-lepton measurements
at STAR in lower energy Au+Au collisions and with detector
upgrades are discussed.

\section{Recent results on di-electron production}\label{results}
We utilize 107 million, 270 million and 150 million events for
$p+p$, minimum-bias (0-80\%) Au+Au, and central (0-10\%) Au+Au
di-electron analyses, respectively. The $p+p$ events were taken in
2009 when 72\% of the full TOF system was installed and
operational, while the Au+Au events were taken in 2010 with full
TOF system coverage. By applying velocity and dE/dx cuts on tracks
with $p_T\!>\!0.2$ GeV/$c$, we can achieve the purity for the
electron candidates of about 99\% in $p+p$ collisions and 97\% in
minimum-bias Au+Au collisions.

The di-electron signals may come from light and heavy flavor
hadron decays. They include $\pi^{0}$, $\eta$, and $\eta^{\prime}$
Dalitz decays: $\pi^{0}\rightarrow \gamma e^{+}e^{-}$, $\eta
\rightarrow \gamma e^{+}e^{-}$, and $\eta^{\prime}\rightarrow
\gamma e^{+}e^{-}$; vector meson decays: $\omega \rightarrow
\pi^{0} e^{+}e^{-}$, $\omega \rightarrow e^{+}e^{-}$, $\rho^{0}
\rightarrow e^{+}e^{-}$, $\phi \rightarrow \eta e^{+}e^{-}$, $\phi
\rightarrow e^{+}e^{-}$, and $J/\psi \rightarrow e^{+}e^{-}$;
heavy-flavor hadron semi-leptonic decays: $c\bar{c} \rightarrow
e^{+}e^{-}X$ and $b\bar{b} \rightarrow e^{+}e^{-}X$; and Drell-Yan
contributions. In Au+Au collisions, we look for additional vector
meson in-medium modifications in the low mass region and possible
QGP thermal radiations in the intermediate mass range.

The $e^{+}$ and $e^{-}$ pairs from the same events are combined to
reconstruct the invariant mass distributions ($M_{ee}$) marked as
unlike-sign distributions. The unlike-sign distributions contain
both signal and background. The background contains random
combinatorial pairs and correlated pairs. The electron candidates
are required to be in the range of $|\eta|\!<1$ and $p_T>0.2$
GeV/$c$ while $e^{+}e^{-}$ pairs are required to be in the
rapidity range of $|y_{ee}|\!<\!1$. Two methods are used for
background estimation, based on same-event like-sign and
mixed-event unlike-sign techniques. In the like-sign technique,
electron pairs with the same charge sign are combined from the
same events. In the mixed-event technique, unlike-sign pairs are
formed from different events. In $p+p$ collisions, we subtract the
like-sign background at $M_{ee}\!<\!0.4$ GeV/$c^{2}$ and
mixed-event background in the higher-mass region. In Au+Au
collisions, we subtract the like-sign background at
$M_{ee}\!<\!0.7$ GeV/$c^{2}$ and mixed-event background in the
higher-mass region. The detailed analysis procedures including
background subtraction and systematic uncertainty evaluations are
published for p+p results in~\cite{ppdilepton:12} and presented
for Au+Au results in~\cite{Jie:11,Jie:12}.

\renewcommand{\floatpagefraction}{0.75}
\begin{figure}[htbp]
\begin{center}
\includegraphics[keepaspectratio,width=0.45\textwidth]{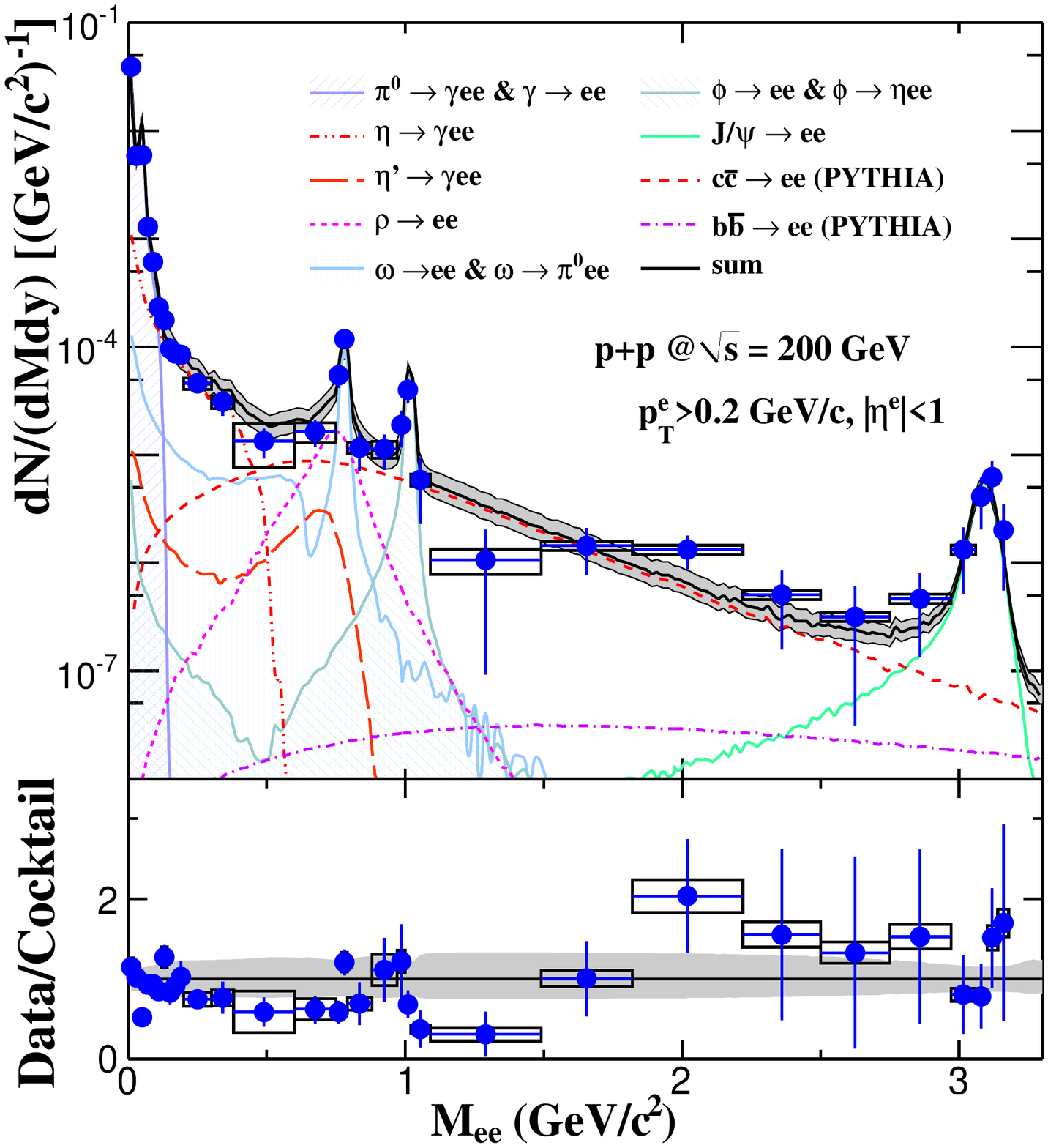}
\includegraphics[keepaspectratio,width=0.45\textwidth]{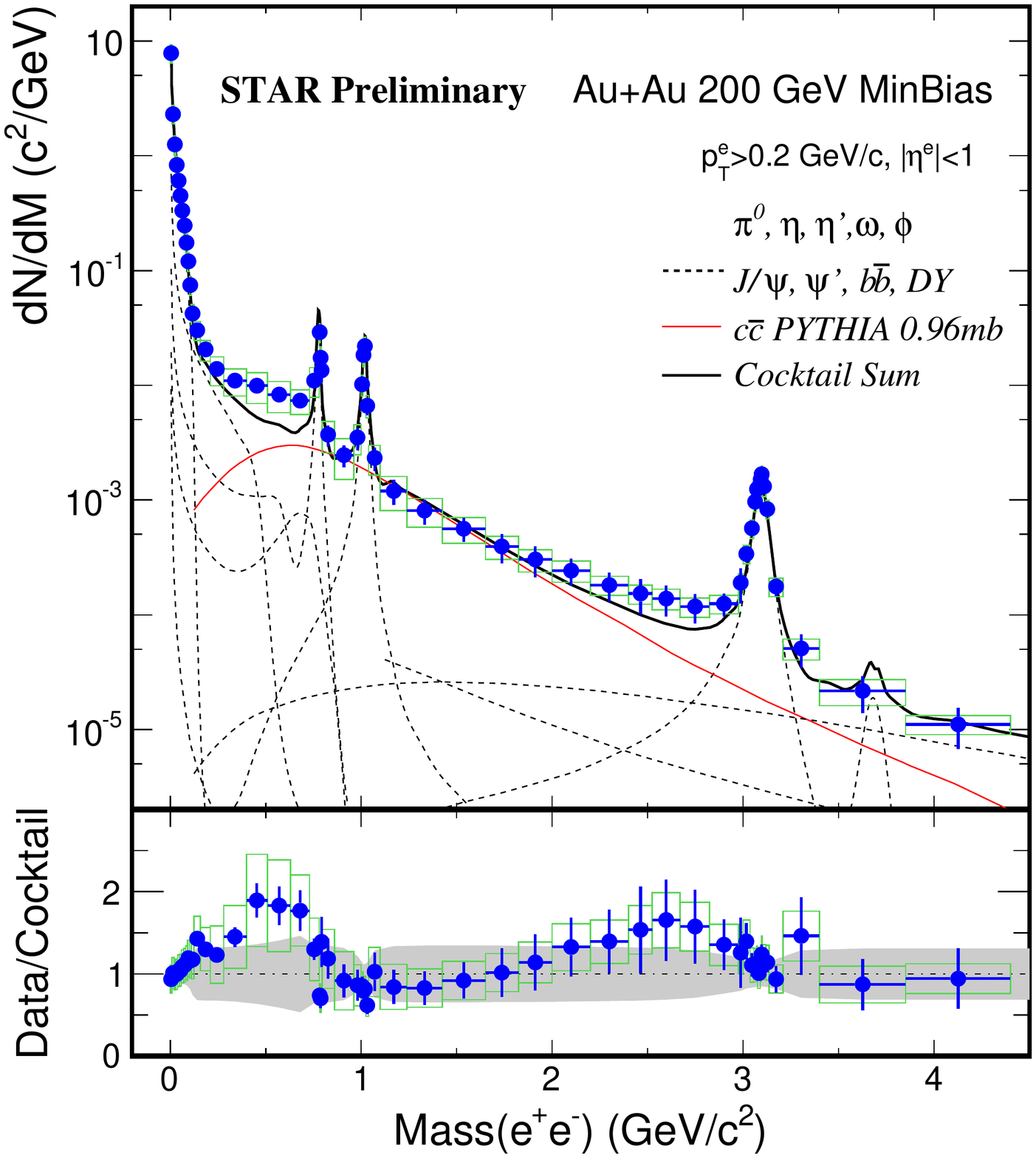}
\includegraphics[keepaspectratio,width=0.45\textwidth]{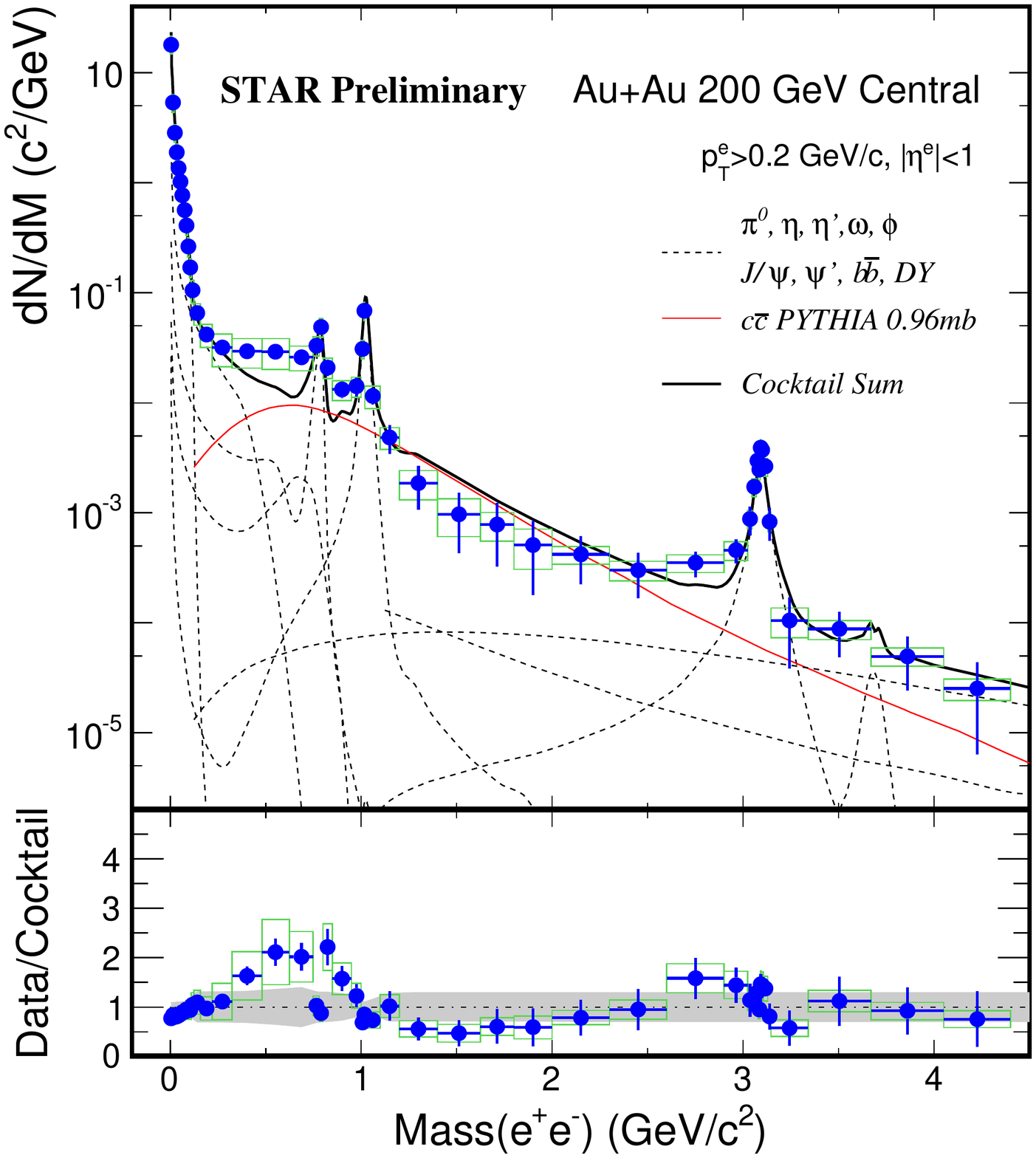}
\begin{minipage}[t]{0.99\textwidth}\caption{(Color online) The comparison for
di-electron continuum between data and simulation after efficiency
correction within the STAR acceptance in $p+p$ (upper-left panel),
minimum-bias (upper-right panel) Au+Au and central (bottom panel)
Au+Au collisions at $\sqrt{s_{_{NN}}} = 200$ GeV. The di-electron
continuum from simulations with different source contributions are
also shown. The Drell-Yan contribution in p+p collisions is not
shown but included in the total cocktail contribution. The bars
and boxes (bands) represent statistical and systematic
uncertainties, respectively. The bands on the bottom panels
illustrate the systematic uncertainties on the cocktail
simulation.} \label{Fig:1}
\end{minipage}
\end{center}
\end{figure}
After the efficiency correction, the di-electron mass spectra
within the STAR acceptance are shown in Fig.~\ref{Fig:1} for
$p+p$, minimum-bias Au+Au and central Au+Au collisions at
$\sqrt{s_{_{NN}}} = 200$ GeV. The di-electron mass spectra are not
corrected for momentum resolution and radiation energy loss
effect. The ratios of data to cocktail simulations are shown in
the lower panels. In $p+p$ collisions, the cocktail simulation,
which includes the expected components from light and heavy flavor
meson decays, is consistent with the measured di-electron
continuum within uncertainties~\cite{ppdilepton:12}. The
$\chi^2/NDF$ between data and cocktail simulation are 21/26 for
$M_{ee}\!>\!0.1$ GeV/$c^2$ and 8/7 for $1.1\!<\!M_{ee}\!<\!3.0$
GeV/$c^2$. In the mass region $0.2\!<\!M_{ee}\!<\!0.8$ GeV/$c^2$,
the cocktail simulation is systematically higher than the measured
di-electron continuum. However, they are also consistent within
uncertainties. We find that better agreement between the cocktail
simulation and data can be achieved by applying an additional
scale factor (56\%) to the $\eta$ Dalitz decay contribution.
Further details on the decay and cocktail simulations are
published in ~\cite{ppdilepton:12}. We also find that the
$c\bar{c} \rightarrow e^{+}e^{-}X$ contribution is dominant in the
intermediate mass region in p+p collisions. In Au+Au collisions,
the $\rho^{0}$ contribution is not included and the $c\bar{c}
\rightarrow e^{+}e^{-}X$ contribution is from PYTHIA
simulation~\cite{pythia} with the total charm cross section 0.96
mb, scaled by the number of underlying binary nucleon-nucleon
collisions.  In the low mass region $0.15\!<M_{ee}\!<\!0.75$
GeV/$c^{2}$, the possible enhancement factors, the ratios of the
data to the cocktail simulations, are $1.53\pm0.07\pm0.41$ and
$1.72\pm0.10\pm0.50$ in minimum-bias and central collisions,
respectively. This suggests for possible vector meson in-medium
modification in this low mass region. The
models~\cite{rapp:09,PSHD:12,USTC:12}, which describe the SPS
di-lepton data but fail to consistently describe the PHENIX
low-$p_T$ and low-mass enhancement~\cite{lowmass}, can describe
our data reasonably well. The comparison between data and models
can be found in~\cite{Jie:12,PSHD:12,USTC:12}. Differential
measurements as a function of $p_T$ and centrality are on-going.

\begin{figure}[htbp]
\begin{center}
\includegraphics[keepaspectratio,width=0.45\textwidth]{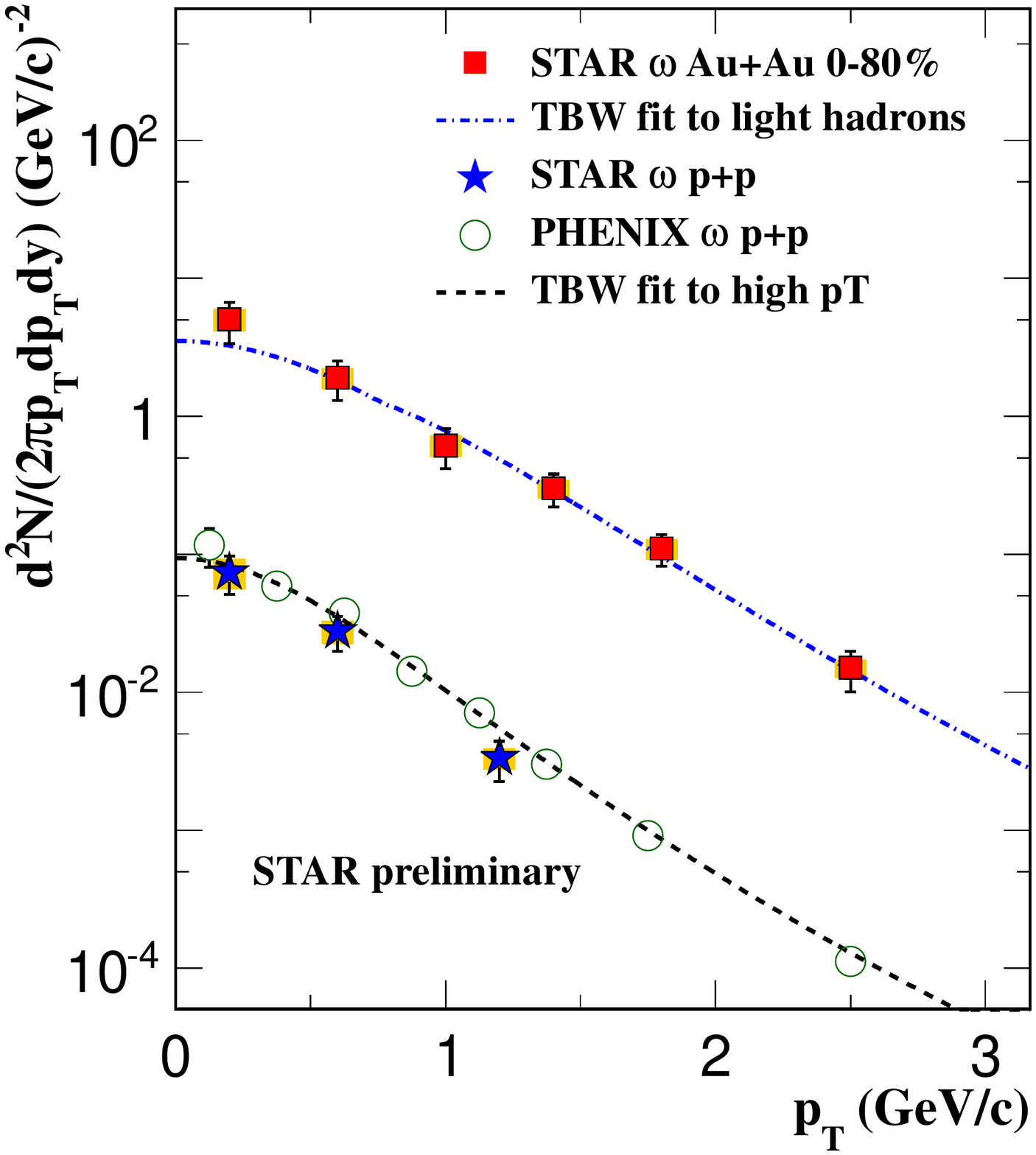}
\includegraphics[keepaspectratio,width=0.45\textwidth]{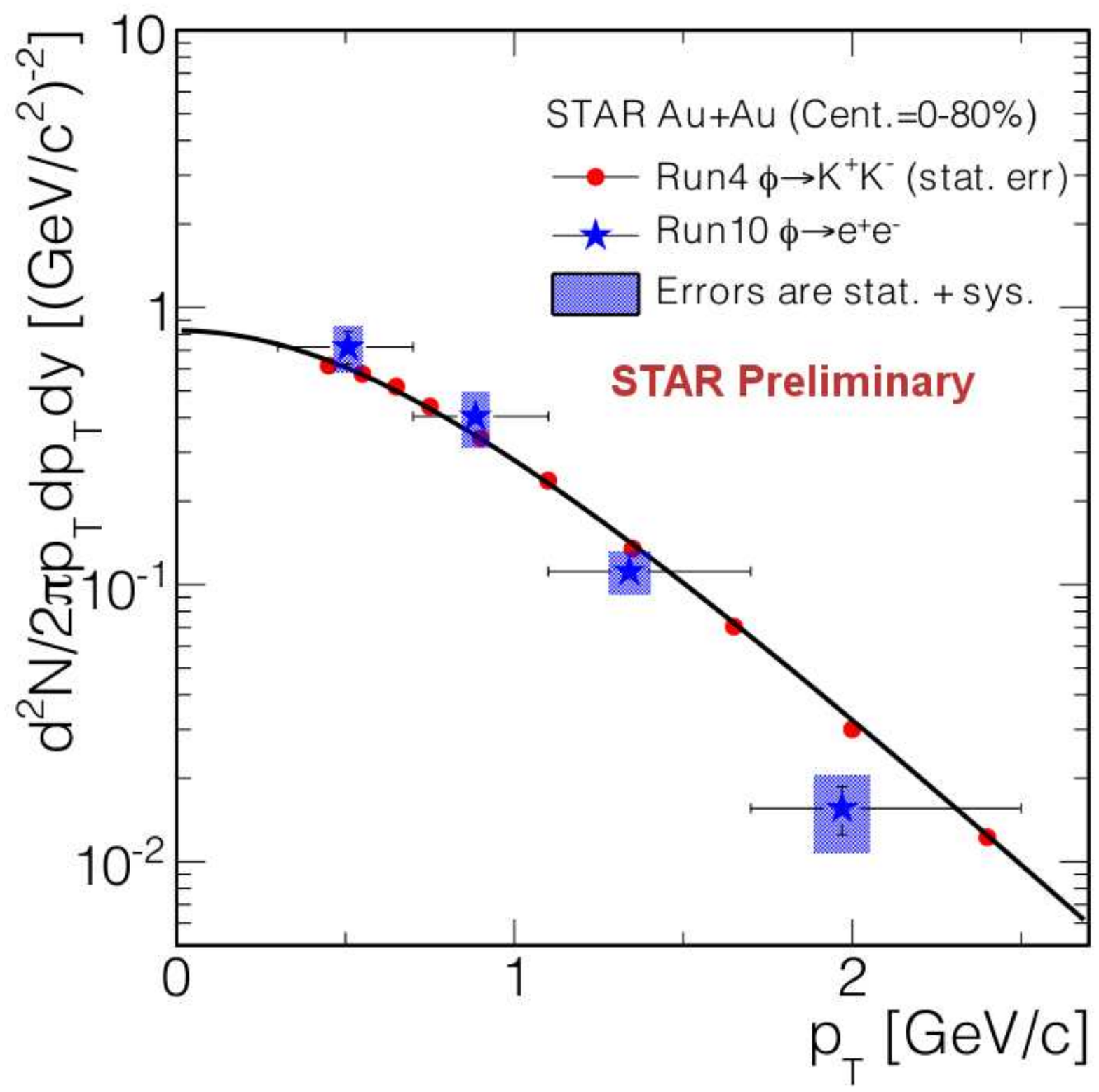}
\begin{minipage}[t]{0.99\textwidth}
\caption{(Color online) (left panel) The $\omega$ invariant yield
measured through di-electronic decay as a function of $p_T$ at
mid-rapidity ($|y|\!<1$) in $p+p$ and minimum-bias Au+Au
collisions at $\sqrt{s_{_{NN}}} = 200$ GeV. The open circles
represent PHENIX published results~\cite{phenixomegappII}. The
dashed line represents the yields of $\omega$ from Tsallis
function (TBW) fit to other particles and high $p_T$ $\omega$
yields in $p+p$ collisions at $\sqrt{s} = 200$ GeV. The dot-dashed
line represents the yields of $\omega$ from Tsallis fit to other
particles in Au+Au collisions at $\sqrt{s_{_{NN}}} = 200$ GeV. The
bars and boxes represent statistical and systematic uncertainties,
respectively. (right panel) The $\phi$ invariant yield measured
through di-electronic and $K^{+}K^{-}$ decays as a function of
$p_T$ at mid-rapidity ($|y|\!<1$) in minimum-bias Au+Au collisions
at $\sqrt{s_{_{NN}}} = 200$ GeV. The curve represents an
exponential fit to the $\phi \rightarrow K^{+}K^{-}$ data points.
The bars represent statistical errors. The boxes represent the
quadrature sum of statistical and systematic uncertainties.}
\label{Fig:2}
\end{minipage}
\end{center}
\end{figure}

In the low mass region, we obtain the invariant yields of $\omega$
and $\phi$ through di-electronic decays at mid-rapidity
($|y|\!<1$) in $p+p$ and Au+Au collisions at $\sqrt{s_{_{NN}}} =
200$ GeV, shown in Fig.~\ref{Fig:2}. In $p+p$ collisions, our
$\omega$ yields from di-electron decays~\cite{ppdilepton:12} are
consistent with previous results~\cite{phenixomegappII} and a
prediction from a Tsallis fit, which fits spectra of other
particles and high $p_T$ ($p_T\!>\!2$ GeV/$c$) $\omega$
yields~\cite{Tsallis,phenixomegapp}. In Au+Au collisions, we fit
light hadrons simultaneously using Tsallis
function~\cite{Tsallis}, obtain the freeze-out parameters,
predict the shape of the $\omega$ invariant yield versus $p_T$,
and find it describes our measurement $\omega \rightarrow
e^{+}e^{-}$~\cite{Bingchu:SQM2011} reasonably well. This indicates
that the $\omega \rightarrow e^{+}e^{-}$ flow pattern is similar
to that of light hadrons in Au+Au collisions. Shown in
Fig.~\ref{Fig:2} (right panel) is the comparison of the invariant
yields of $\phi$ through its $e^{+}e^{-}$~\cite{Masa:SQM2011} and
$K^{+}K^{-}$~\cite{phi:09} decays. Within statistical and
systematic uncertainties, the $\phi$ invariant yields measured
through $e^{+}e^{-}$ and $K^{+}K^{-}$ decays are consistent.
\renewcommand{\floatpagefraction}{0.75}

\begin{figure}[htbp]
\begin{center}
\includegraphics[keepaspectratio,width=0.45\textwidth]{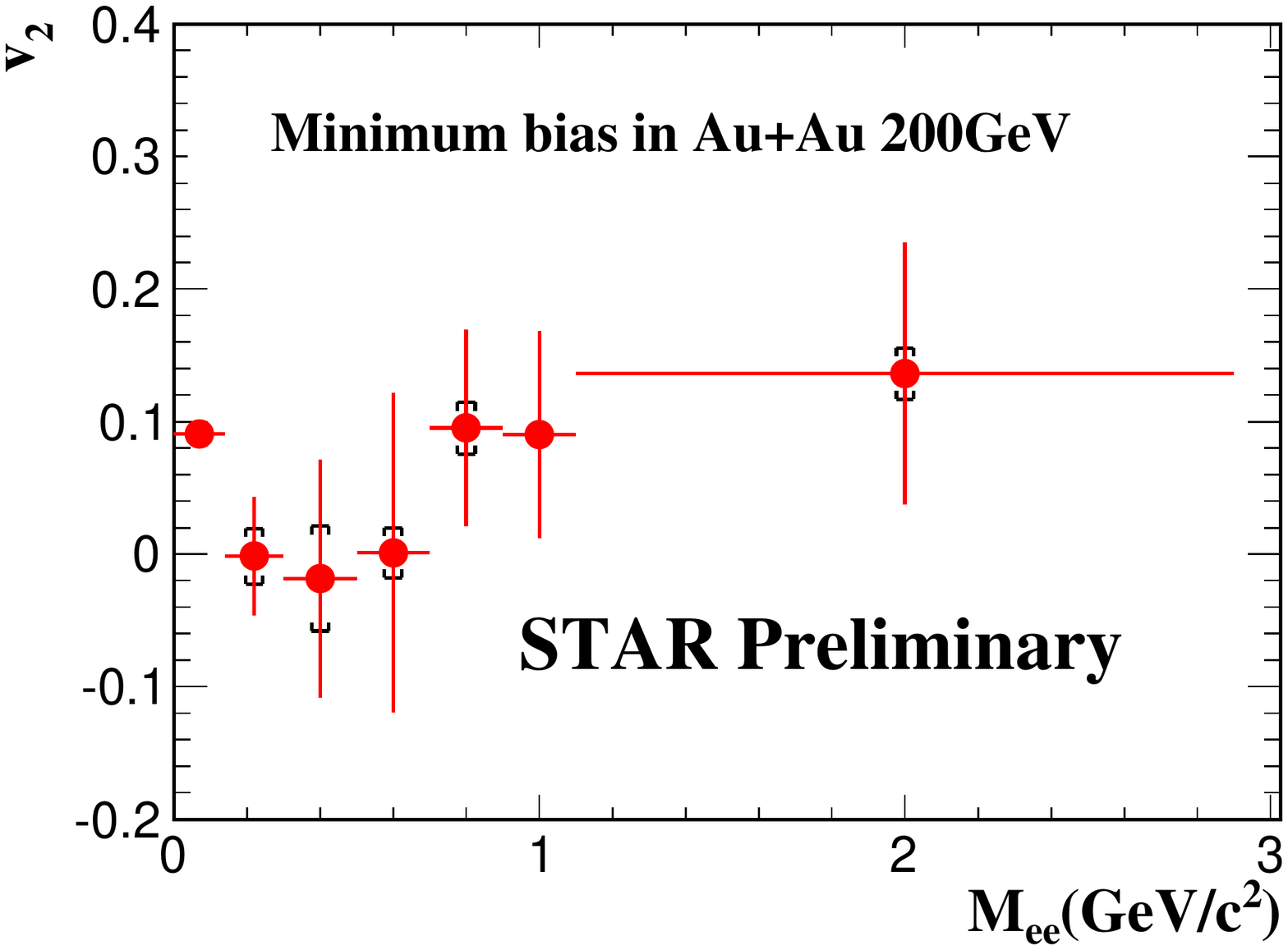}
\includegraphics[keepaspectratio,width=0.45\textwidth]{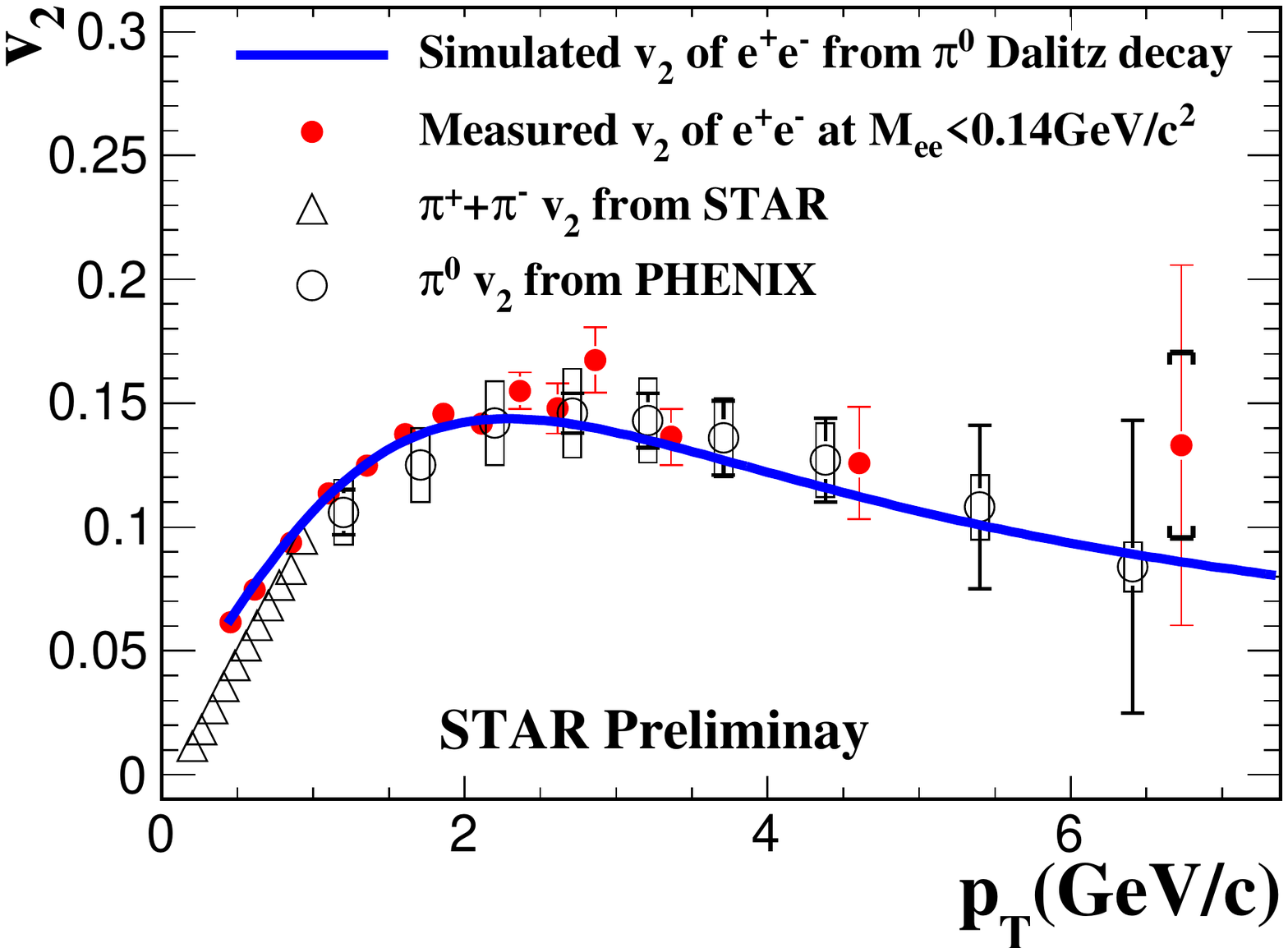}
\includegraphics[keepaspectratio,width=0.45\textwidth]{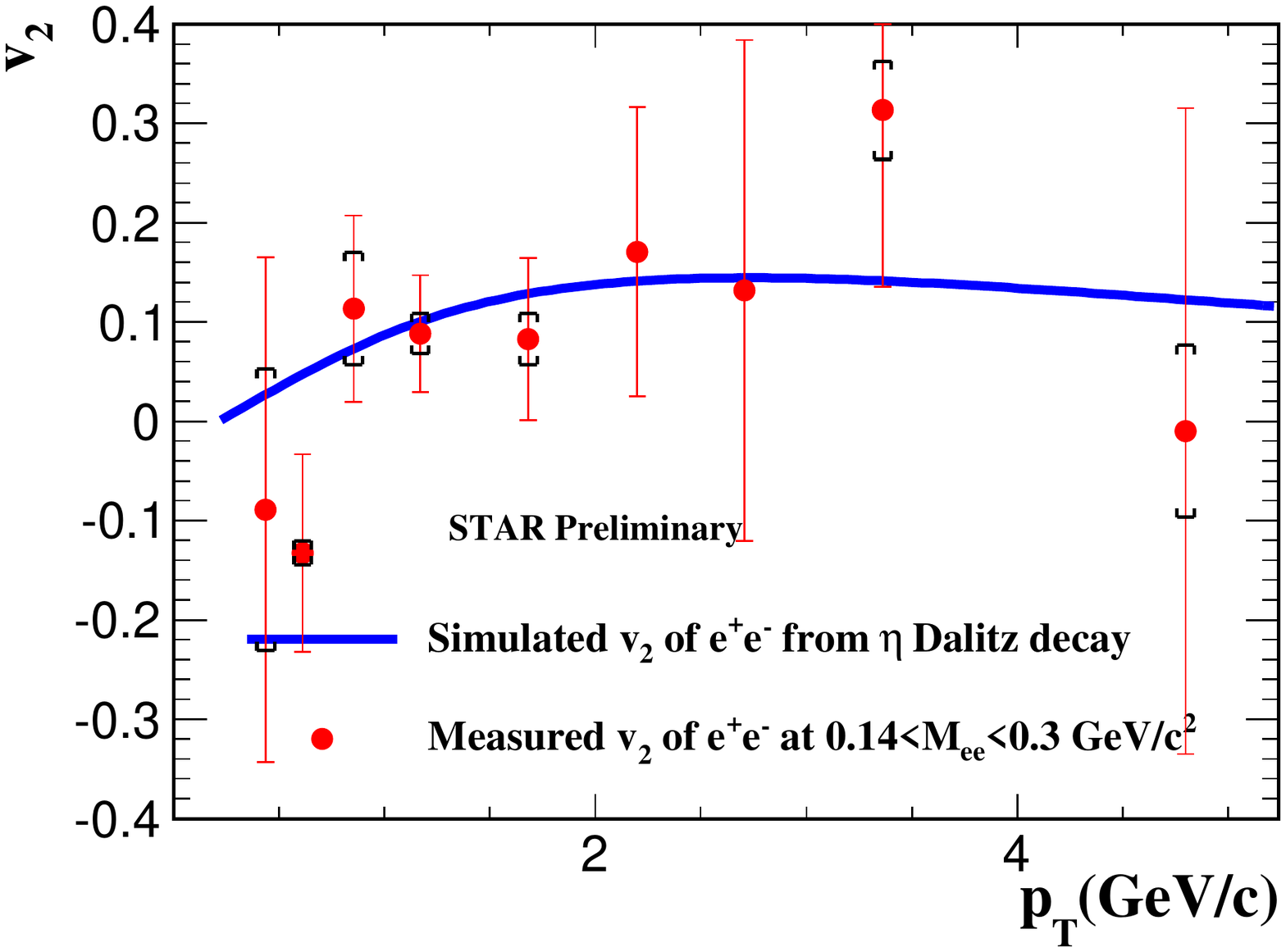}
\begin{minipage}[t]{0.99\textwidth}
\caption{(Color online) (upper-left panel) The di-electron $v_2$
as a function of $M_{ee}$ in minimum-bias Au+Au collisions at
$\sqrt{s_{_{NN}}} = 200$ GeV. (upper-right panel) The $v_2$ of
di-electron at $M_{ee}\!<\!0.14$ GeV/$c^{2}$ (solid symbols) as a
function of $p_T$ in minimum-bias Au+Au collisions at
$\sqrt{s_{_{NN}}} = 200$ GeV. Also shown are the charged and
neutral pion $v_2$ and the expected $v_2$ (solid curve) of
di-electrons from $\pi^{0}$ Dalitz decay. (bottom panel) The $v_2$
of di-electron at $0.14\!<M_{ee}\!<\!0.30$ GeV/$c^{2}$ as a
function of $p_T$ in minimum-bias Au+Au collisions at
$\sqrt{s_{_{NN}}} = 200$ GeV. Also shown is the expected $v_2$
(solid curve) of di-electrons from $\eta$ Dalitz decay. The bars
and boxes represent statistical and part of systematic
uncertainties, respectively.} \label{Fig:3}
\end{minipage}
\end{center}
\end{figure}

The di-lepton $v_2$ measurements provide another independent way
to study the medium properties. We use event-plane method to
obtain the di-electron $v_2$. The event-plane is reconstructed
using the tracks from the TPC. The details of the method are in
Refs.~\cite{Art:98,Yan:08}. We report the $v_2$ of di-electron
signals in Fig.~\ref{Fig:3} (upper-left panel) as a function of
$M_{ee}$ in minimum-bias Au+Au collisions at $\sqrt{s_{_{NN}}} =
200$ GeV. The differential $v_2$ of di-electron pairs in the mass
regions of $M_{ee}\!<\!0.14$ GeV/$c^{2}$ and
$0.14\!<M_{ee}\!<\!0.30$ GeV/$c^{2}$ are shown respectively in the
upper-right and bottom panels of Fig.~\ref{Fig:3} as a function of
$p_T$ in minimum-bias Au+Au collisions at $\sqrt{s_{_{NN}}} = 200$
GeV. Also shown are the charged~\cite{YutingThesis} and neutral
pion~\cite{PHENIXv2} $v_2$. The dominant sources to di-electrons
at $M_{ee}\!<\!0.14$ GeV/$c^{2}$ and $0.14\!<M_{ee}\!<\!0.30$
GeV/$c^{2}$ are $\pi^{0}$ Dalitz decay and $\eta$ Dalitz decay,
respectively. We parameterize the pion $v_2$ from low to high
$p_T$~\cite{YutingThesis,PHENIXv2}, perform the Dalitz decay
simulation, and obtain the expected di-electron $v_2$ from $\pi^0$
Dalitz decay shown by the solid curve. The simulated $v_2$ is
consistent with the measured di-electron $v_2$ at
$M_{ee}\!<\!0.14$ GeV/$c^{2}$. The consistency between the
expectations and measurements demonstrates the credibility of our
method to obtain the di-electron $v_2$. We repeat the same
exercise in the $\eta$ mass region. We assume that $\eta$ has the
same $v_2$ as $K_{S}^{0}$~\cite{Yan:08} since the $\eta$ mass is
close to that of $K_{S}^{0}$. The simulated $v_2$ of di-electrons
from $\eta$ Dalitz decay, shown as a solid curve, is consistent
with the measured di-electron $v_2$ at $0.14\!<M_{ee}\!<\!0.30$
GeV/$c^{2}$. The current precision of our $v_2$ data does not
allow to further study a possible deviation from the solid curve
due to the other contributions in this mass region.
\renewcommand{\floatpagefraction}{0.75}

\begin{figure}[htbp]
\begin{center}
\includegraphics[keepaspectratio,width=0.45\textwidth]{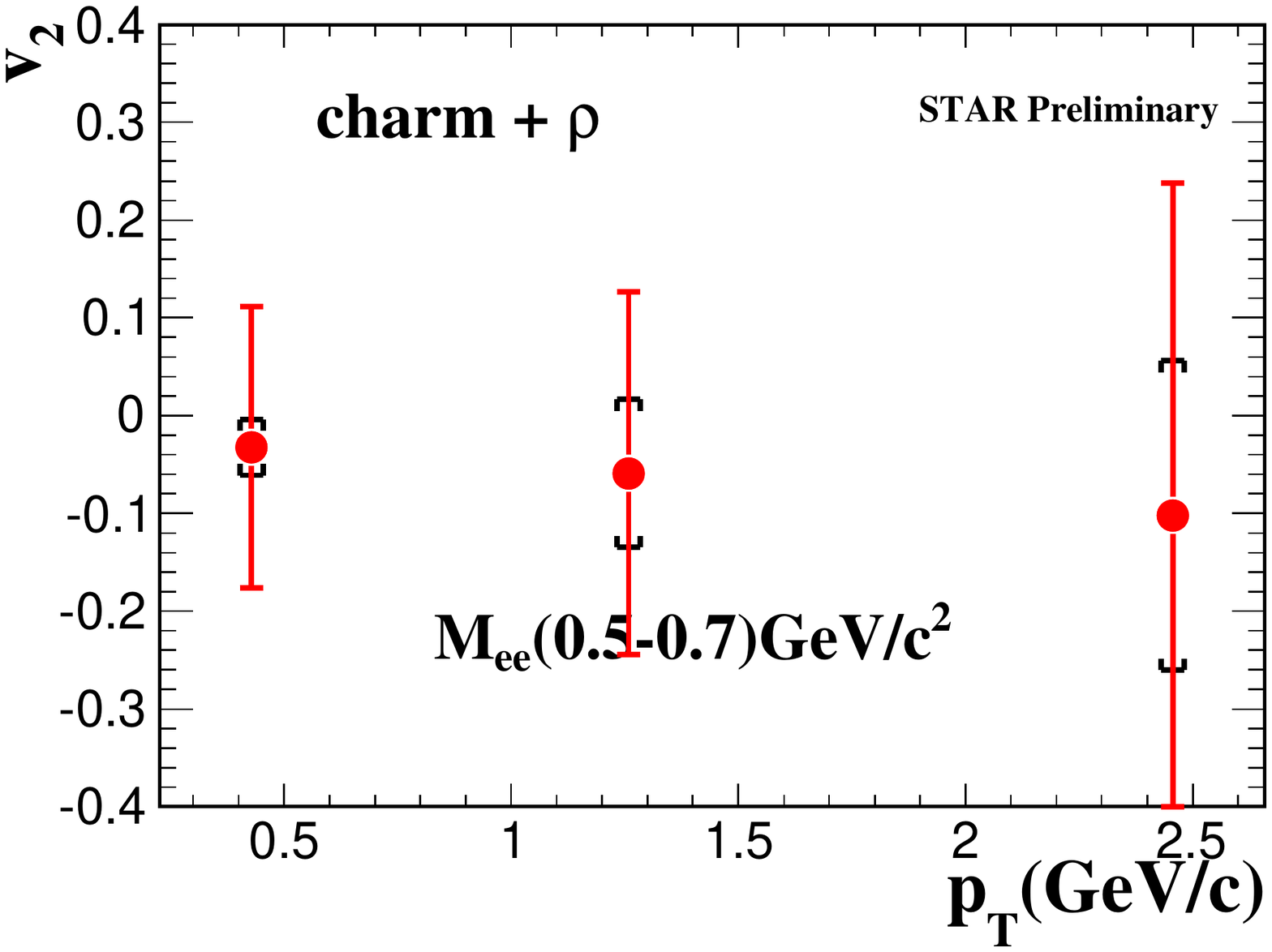}
\includegraphics[keepaspectratio,width=0.45\textwidth]{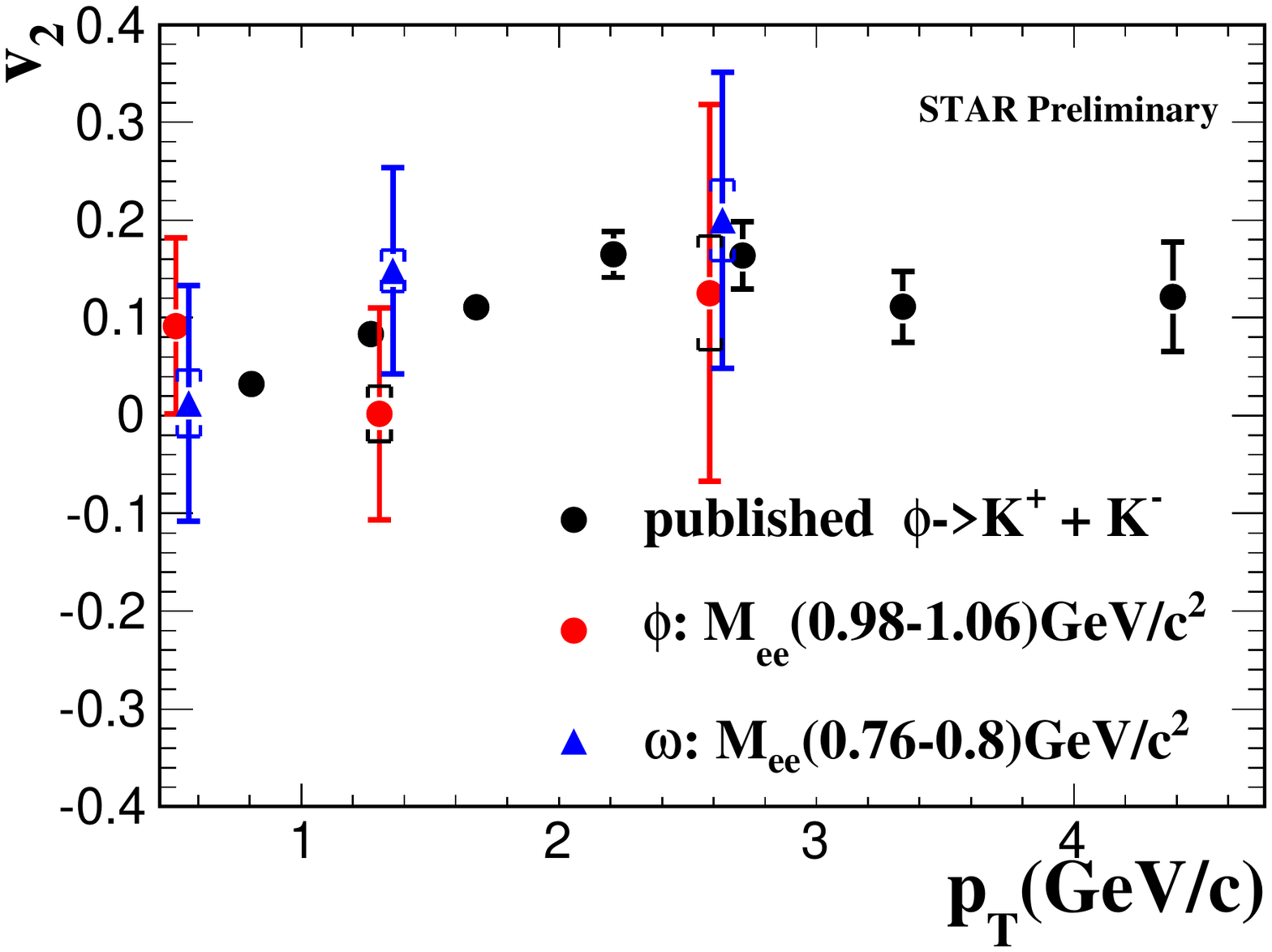}
\begin{minipage}[t]{0.99\textwidth}
\caption{(Color online) The di-electron $v_2$ as a function of
$p_T$ in minimum-bias Au+Au collisions at $\sqrt{s_{_{NN}}} = 200$
GeV for $0.5\!<M_{ee}\!<\!0.7$ GeV/$c^{2}$ (left),
$0.76\!<M_{ee}\!<\!0.8$ GeV/$c^{2}$, and $0.98\!<M_{ee}\!<\!1.06$
GeV/$c^{2}$ (right). Also shown is the measured $v_2$ of $\phi$
meson through the $K^{+}K^{-}$ decay. The bars and boxes represent
statistical and part of systematic uncertainties, respectively.}
\label{Fig:4}
\end{minipage}
\end{center}
\end{figure}

Figure~\ref{Fig:4} (left panel) shows the di-electron $v_2$ as a
function of $p_T$ in minimum-bias Au+Au collisions at
$\sqrt{s_{_{NN}}} = 200$ GeV for $0.5\!<M_{ee}\!<\!0.7$
GeV/$c^{2}$ in which charm correlation and in-medium $\rho$
contribution might be dominant. Right panel of Fig.~\ref{Fig:4}
shows the di-electron $v_2$ as a function of $p_T$ in the mass
ranges of $0.76\!<M_{ee}\!<\!0.8$ GeV/$c^{2}$ and
$0.98\!<M_{ee}\!<\!1.06$ GeV/$c^{2}$ in minimum-bias Au+Au
collisions at $\sqrt{s_{_{NN}}} = 200$ GeV. The di-electrons $v_2$
for $0.98\!<M_{ee}\!<\!1.06$ GeV/$c^{2}$ is consistent with the
measured $v_2$ of $\phi$ meson through the $K^{+}K^{-}$
decay~\cite{phiv2:07}.
\renewcommand{\floatpagefraction}{0.75}

\section{Future perspectives}\label{future}
A factor of two more Au+Au data at $\sqrt{s_{_{NN}}} = 200$ GeV,
taken in 2011, will significantly improve the measurements of mass
spectra and elliptic flow. The possible low mass enhancement
factors for $0.15\!<M_{ee}\!<\!0.75$ GeV/$c^{2}$ are significantly
lower than those measured by PHENIX in minimum-bias and central
collisions~\cite{Jie:12}. The $p_T$ dependence measurements in the
future will allow more differential comparisons between STAR and
PHENIX. In addition, di-electron $v_2$ as a function of $p_T$ in
the intermediate mass region $1.1\!<M_{ee}\!<\!2.9$ GeV/$c^{2}$
will be obtained.

In 2010 and 2011, STAR has taken a few hundred million
minimum-bias events in Au+Au collisions at $\sqrt{s_{_{NN}}} =
19.6, 27, 39,$ and $62.4$ GeV with full TOF azimuthal coverage and
low conversion material budget, which will enable us to
systematically study the energy dependence of the following
physics topics: 1) di-electron enhancement in the low mass
region~\cite{phenixppdilepton,phenixdielectron}; 2) in-medium
modifications of vector meson decays; 3) virtual
photons~\cite{thermalphoton}; 4) $c\bar{c}$ medium modifications;
and 5) possible QGP thermal radiation in the intermediate mass
region. Specifically, the energy value of 19.6 GeV is comparable
to the center of mass energy for the CERES and NA60 measurements.
The di-electron results in Au+Au collisions at $\sqrt{s_{_{NN}}} =
19.6$ GeV will enable another consistency check between STAR
measurements and previous SPS results.

With the current data sets, it will be difficult to measure charm
correlation contribution or QGP thermal radiation in the
intermediate mass region since they are coupled to each other and
one is the other's background for the physics case. So far at
RHIC, there is no clear answer about thermal radiation in the
intermediate mass region. The future detector upgrade with the
Heavy Flavor Tracker at STAR, to be completed in 2014, will
provide precise charm cross section measurements~\cite{hft}. This
will help to understand heavy quark dynamics in the medium and
constrain model inputs to calculate di-leptons from heavy flavor
correlations. However the measurements of $c\bar{c}$ correlations
will still be challenging if not impossible. An independent
approach is proposed with the Muon Telescope Detector upgrade
(MTD)~\cite{starmtdproposal}. The $\mu-e$ correlations measure the
contribution from heavy flavor correlations to the di-electron or
di-muon continuum. This will make it possible to access the
thermal radiation in the intermediate mass region.

\begin{figure}[htbp]
\begin{center}
\includegraphics[keepaspectratio,width=0.45\textwidth]{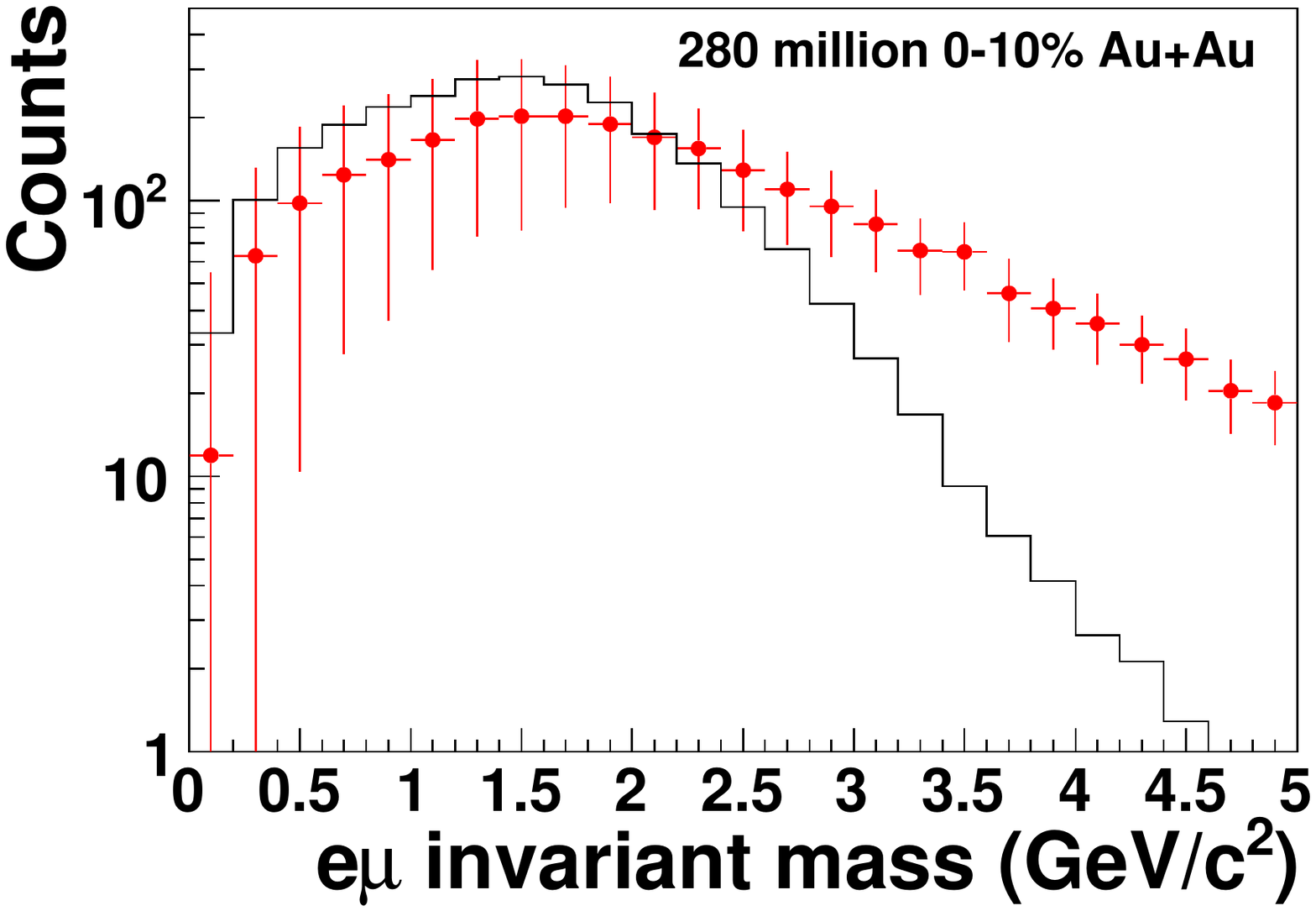}
\includegraphics[keepaspectratio,width=0.45\textwidth]{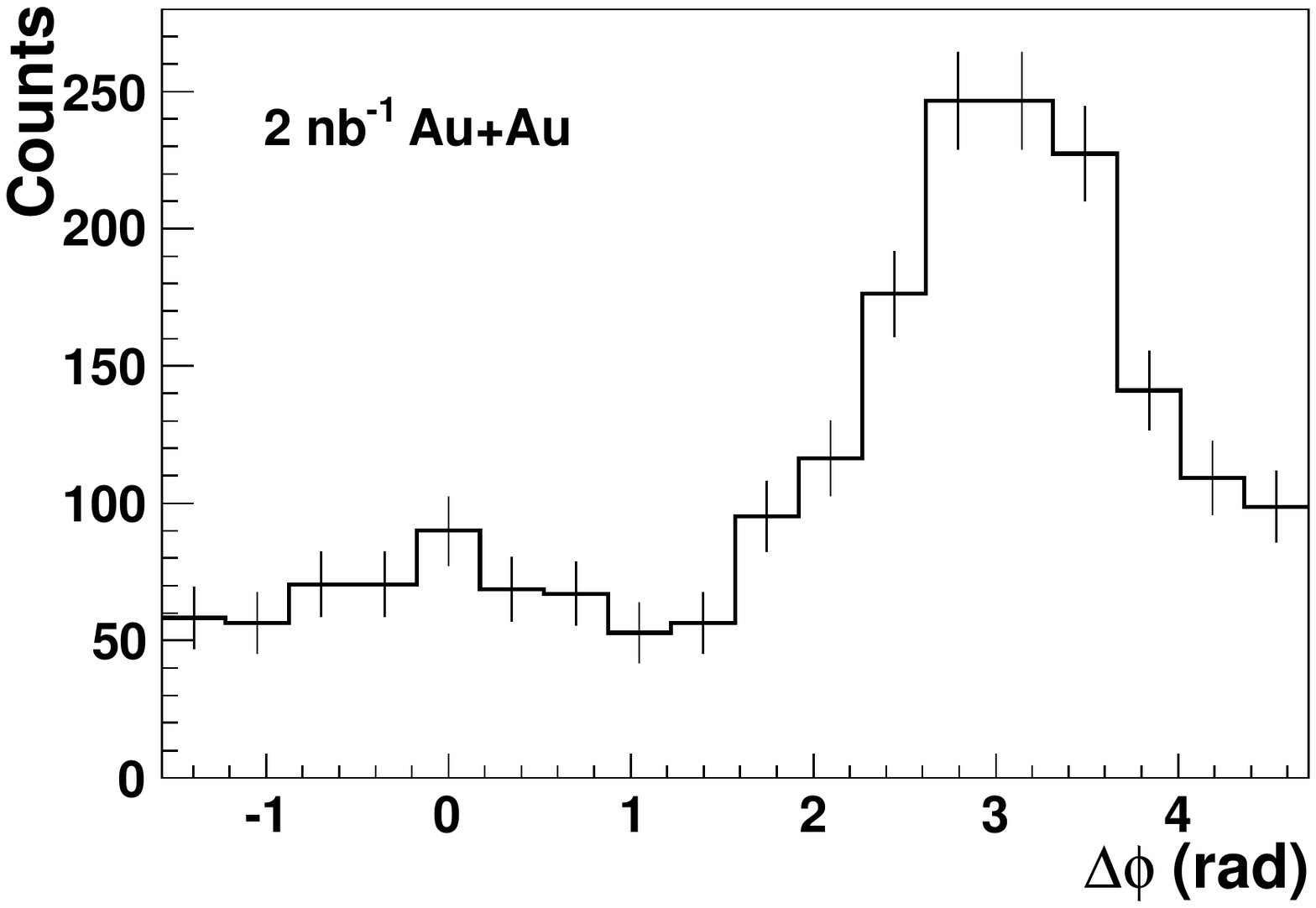}
\begin{minipage}[t]{0.99\textwidth}
\caption{(Color online) (left panel) The precision projection for
$\mu-e$ invariant mass distribution from charm correlation
contribution from simulation in Au+Au collisions at
$\sqrt{s_{_{NN}}} = 200$ GeV for Run 2013. The circles represent
the case that the $\mu-e$ correlation is not modified in Au+Au
collisions for 280 million central events. The black line
represents the case that $c$ and $\bar{c}$ is de-correlated.
(right panel) The precision projection for $\mu-e$ azimuthal
angular correlation from charm correlation contribution from
PYTHIA simulation for Run 2013 when we sample 2 $nb^{-1}$ Au+Au
luminosity with a coincidence trigger of single muon hit and BEMC
energy deposition above a certain threshold. The projection is
done under the assumption that the correlation is the same as in
PYTHIA. } \label{Fig:5}
\end{minipage}
\end{center}
\end{figure}

The MTD construction, to be completed in 2014, has started. In
2012, 10\% of the MTD was installed at STAR and worked nicely with
smooth data taking. This will enable a first proof-of-principle
for $\mu-e$ measurement. In 2013, 43\% of MTD will be installed
and we request three-week Au+Au run at 200 GeV for the $\mu-e$
measurement. Figure~\ref{Fig:5} shows the precision projections
for $\mu-e$ invariant mass distribution and azimuthal angular
correlation from charm correlation contribution from simulation in
Au+Au collisions at $\sqrt{s_{_{NN}}} = 200$ GeV for Run 2013. If
the $\mu-e$ correlation is not modified in Au+Au collisions, the
precision projection is indicated by the circle shown in
Fig.~\ref{Fig:5} (left panel) for 280 million central events. Also
shown for the comparison is the case that $c$ and $\bar{c}$ is
de-correlated illustrated by the black line. In addition, with
full azimuthal coverage of the TPC, TOF, and Barrel
Electro-magnetic Calorimeter (BEMC), we will be able to measure
$\mu-e$ azimuthal angular correlation. Figure~\ref{Fig:5} (right
panel) illustrates the precision projection for $\mu-e$ azimuthal
angular correlation from PYTHIA simulation when we sample 2
$nb^{-1}$ Au+Au luminosity with a coincidence trigger of single
muon hit and BEMC energy deposition above a certain threshold. The
projection is done under the assumption that the correlation is
the same as in PYTHIA.

\renewcommand{\floatpagefraction}{0.75}

\section{Summary}\label{summary}
In summary, the di-electron mass spectra are measured in 200 GeV
$p+p$ and Au+Au collisions at STAR. The cocktail simulations are
consistent with the data in 200 GeV $p+p$ collisions. In Au+Au
collisions, we observe a possible enhancement by comparison
between data and cocktail simulation in the low mass region
$0.15\!<M_{ee}\!<\!0.75$ GeV/$c^{2}$. The first elliptic flow
measurements of di-electrons are presented in 200 GeV minimum-bias
Au+Au collisions. The $v_2$ of di-electrons at $M_{ee}\!<\!0.14$
GeV/$c^{2}$ and $0.14\!<M_{ee}\!<\!0.30$ GeV/$c^{2}$ are in
agreement with the expectations from previous measurements. In the
future, more precise differential measurements will be obtained
for di-electron spectra and $v_2$ at 200 GeV. The data taken at
lower energies will allow to systematically study the energy
dependence of low-mass enhancement. The precise charmed hadron
measurements from future detector upgrade with the Heavy Flavor
Tracker will help to constrain model inputs to calculate
di-leptons from heavy flavor correlations. The $\mu-e$
correlations with the Muon Telescope Detector upgrade will measure
the contribution from heavy flavor correlations to the di-electron
or di-muon continuum. This will make it possible to access the
thermal radiation in the intermediate mass region.












\end{document}